\documentclass{PoS}
\makeatletter\let\default@color\current@color\makeatother 

\usepackage{amsmath,amssymb,amsbsy}

\def\CO{{\cal O}}

\def\spose#1{\hbox to 0pt{#1\hss}}
\def\ltapprox{\mathrel{\spose{\lower 3pt\hbox{$\mathchar"218$}}
 \raise 2.0pt\hbox{$\mathchar"13C$}}}
\def\gtapprox{\mathrel{\spose{\lower 3pt\hbox{$\mathchar"218$}}
 \raise 2.0pt\hbox{$\mathchar"13E$}}}
\def\inapprox{\mathrel{\spose{\lower 3pt\hbox{$\mathchar"218$}}
 \raise 2.0pt\hbox{$\mathchar"232$}}}

\title{The kaon B-parameter from unquenched mixed action lattice QCD}

\ShortTitle{$B_K$ from unquenched mixed action lattice QCD}

\author{Christopher Aubin\\
        Department of Physics, Columbia University, New York, NY, USA\\
        Department of Physics, College of William and Mary,\thanks{Present address.}~~Williamsburg, VA, USA\\
        E-mail: \email{caaubin@wm.edu}}
        
\author{Jack Laiho\\
        Theoretical Physics Department,
        Fermi National Accelerator Laboratory,
        Batavia, Illinois, USA \\
        Department of Physics,
        Washington University,\thanks{Present address.}~~St.\ Louis, Missouri, USA \\
        E-mail: \email{jlaiho@fnal.gov}}
        
\author{\speaker{Ruth S. Van de Water}\\
        Theoretical Physics Department,
        Fermi National Accelerator Laboratory,\thanks{Operated by Fermi  Research Alliance, LLC,
         under Contract No.~DE-AC02-07CH11359 with the  United States Department of Energy.}~~Batavia, Illinois, USA \\
        E-mail: \email{ruthv@fnal.gov}}

\abstract{We present a preliminary calculation of $B_K$ using domain-wall valence quarks and 2+1 flavors of improved staggered sea quarks.  Both the size of the residual quark mass, which measures the amount of chiral symmetry breaking, and of the mixed meson splitting $\Delta_\textrm{mix}$, a measure of taste-symmetry breaking, show that discretization effects are under control in our mixed action lattice simulations.  We show preliminary data for pseudoscalar meson masses, decay constants and $B_K$.  We discuss general issues associated with the chiral extrapolation of lattice data, and, as an example, present a preliminary chiral and continuum extrapolation of $f_\pi$.  The quality of our data shows that the good chiral properties of domain-wall quarks, in combination with the light sea quark masses and multiple lattice spacings available with the MILC staggered configurations, will allow for a precise determination of $B_K$.
}

\FullConference{The XXV International Symposium on Lattice Field Theory\\
		 July 30 - August 4 2007\\
		 Regensburg, Germany}

\begin{document}

\section{Lattice calculation of $B_K$ with a mixed action}

The kaon B-parameter ($B_K$), which parameterizes the hadronic part of neutral kaon mixing, plays an important role in flavor physics phenomenology.  When combined with an experimental measurement of $\epsilon_K$, $B_K$ constrains the apex of the  Cabibbo-Kobayashi-Maskawa unitarity triangle.  Because $\epsilon_K$ is well-known, the dominant uncertainty in this constraint is that of $B_K$.\footnote{While $B_K$ is currently the dominant source of uncertainty, the importance of other quantities such as the perturbative Inami-Lim functions and $|V_{cb}|$ will increase as the precision on $B_K$ improves.}  It is likely that whatever new physics exists has additional $CP$-violating phases;  these will manifest themselves as inconsistencies between measurements that are predicted to be identical within the Standard Model.  Thus a precise determination of $B_K$ will help constrain physics beyond the Standard Model.

The determination of $B_K$ is an important goal of the lattice QCD community.  Thus many calculations have been done, each improving upon the previous one.  The benchmark JLQCD calculation involved a thorough study of the quark mass and lattice spacing dependence, but did not include the effect of sea quark loops, resulting in an indeterminate quenching error~\cite{JLQCD_BK}.  The HPQCD collaboration eliminated this uncertainty by using dynamical staggered fermions, but their result has a $\sim 20\%$ systematic error from neglected higher-order and wrong-taste operators in the lattice-to-continuum matching procedure~\cite{HPQCD_BK}. The RBC and UKQCD Collaborations recently calculated $B_K$ with dynamical domain-wall fermions and fully nonperturbative operator renormalization, but they have only a single lattice spacing and cannot yet perform a continuum extrapolation~\cite{RBC_BK}.

Our mixed action calculation combines domain-wall valence quarks and staggered sea quarks.   We use the MILC 2+1 flavor improved staggered lattices which are publicly available with a large range of quark masses, lattice spacings, and volumes and allow for good control over the systematic error from chiral and continuum extrapolation~\cite{bigMILC}.  We use the domain-wall propagator code from the Chroma lattice QCD software package~\cite{Chroma}.  Because domain-wall quarks do not carry taste quantum numbers, the $B_K$ lattice operator mixes only with other operators of incorrect chirality, making the chiral extrapolation more continuum-like than in the purely staggered case, and allowing for nonperturbative operator matching using the Rome-Southampton method~\cite{NPR}.  Thus the mixed action method combines the advantages of staggered and domain-wall fermions without suffering from their primary disadvantages and is well-suited to the calculation of $B_K$.

\section{Discretization effects in mixed action lattice simulations}

Each flavor of staggered quark comes in four identical species, or ``tastes", that are related by an $SU(4)$ symmetry in the continuum~\cite{Susskind}.  The staggered taste symmetry, however, is broken at nonzero lattice spacing, causing the masses of the sixteen tastes of sea-sea mesons in mixed action simulations to be split according to their $SO(4)$ taste representation at $\CO(a^2)$~\cite{LeeSharpe}:
\begin{equation}
	m^2_{SS',t} = \mu (m_S + m_{S'}) + a^2 \Delta_t .
\end{equation}
These taste-splittings generically enter staggered $\chi$PT expressions for masses and weak matrix elements.  They have already been measured, however, for the MILC lattices in Ref.~\cite{bigMILC}, and we use these values in the chiral and continuum extrapolation of our lattice data.

\begin{figure}
\begin{center}
\includegraphics[scale=0.37,angle=0]{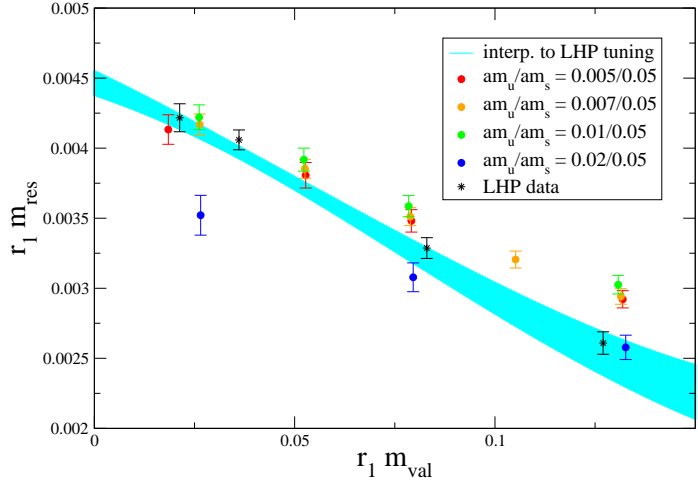}\end{center}
\vspace{-3mm}
\caption{Chiral extrapolation of $m_{res}$ on the coarse MILC lattices.  Because there is no notion of full QCD at nonzero lattice spacing in the mixed action theory, the curve shows the extrapolation/interpolation for points where the domain-wall pion mass is tuned to equal the lightest (taste pseudoscalar) staggered pion mass.  For comparison, we show the determination of $m_{res}$ by the LHP collaboration, which uses this tuning~\cite{LHPC}.}
\label{fig:mres}
\end{figure}

Domain-wall quarks receive an additive contribution to their mass from explicit chiral symmetry breaking, the size of which is controlled by the length of the fifth dimension~\cite{DWF}.  Consequently, to lowest order in $\chi$PT, the masses of valence-valence mesons in the mixed action theory are
 \begin{equation}
 	m^2_{VV'} = \mu (m_V + m_{V'} + 2 m_{res}),
 \end{equation}
where $m_{res}$ is the residual quark mass.  In our simulations the length of the fifth dimension is $L_S=16$.  With this choice, we determine a preliminary value of $r_1 m_{res} = 0.00446(9)(45)$, around 3~MeV, in the chiral limit; the extrapolation is shown in Figure~\ref{fig:mres}.  
Because $m_{res}$ is a quarter the size of our lightest valence quark mass, and half that of RBC/UKQCD (in physical units), this indicates that chiral symmetry breaking is under control in our mixed action lattice simulations.  

\begin{figure}
\begin{center}
\includegraphics[scale=0.43,angle=0]{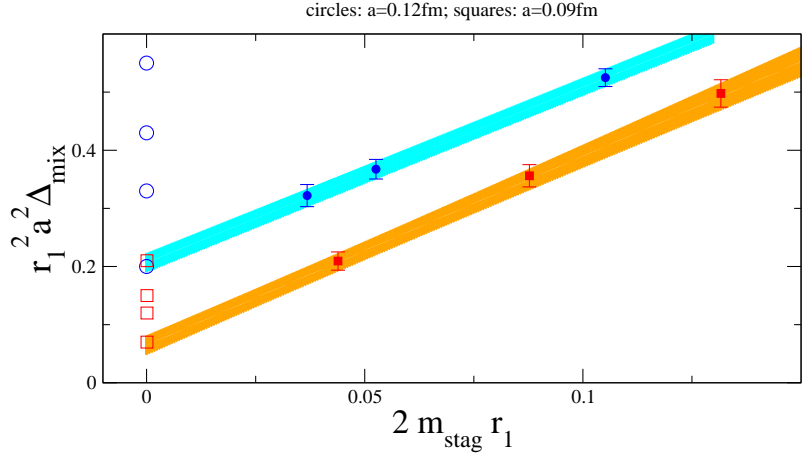} \end{center}
\vspace{-3mm}
\caption{Determination of $\Delta_\textrm{mix}$ on the MILC coarse (circles) and fine (squares) lattices.  The filled symbols are our data points.  For comparison, the open symbols show the four MILC staggered taste splittings.}
\label{fig:DelMix}
\end{figure}

Because the mixed action lattice theory has new four-fermion operators, the chiral effective theory can have new low-energy constants.  It turns out, however, that the mixed action chiral Lagrangian has only one new constant at lowest order~\cite{Bar}.  This coefficient, due to taste-symmetry breaking from the sea quarks, produces an $\CO(a^2)$ shift to the mixed valence-sea meson mass-squared:
\begin{equation}
	m^2_{VS} = \mu ( m_V + m_{res} + m_S) + a^2 \Delta_\textrm{mix} .
\end{equation}
In order to determine $\Delta_\textrm{mix}$ for our lattice simulations, we construct a linear combination of valence-valence and valence-sea squared meson masses,
\begin{equation}
\frac{1}{2}(2 m^2_{VV} - m^2_{VS}) = a^2 \Delta_\textrm{mix} + 2 \mu m_{S} ,
\end{equation}
and perform a linear fit versus the staggered quark mass; this fit is shown in Figure~\ref{fig:DelMix}.  We obtain $r_1^2 a^2 \Delta_\textrm{mix} = 0.206 (16)(21)$, where the systematic error is preliminary.  This is close to the staggered sea splitting $\Delta_A$.  It is also consistent with the independent determination of $\Delta_\textrm{mix}$ by Orginos \& Walker-Loud~\cite{MixMeson}.  Because we have data at two lattice spacings, we can also determine the scaling behavior of $\Delta_\textrm{mix}$.  We find that it scales correctly for the MILC lattices, as $\CO(\alpha_S^2 a^2)$.

\section{Preliminary data and analysis}

We have generated quark propagators on the MILC coarse ($a \approx 0.12$ fm) and fine ($a \approx 0.09$ fm) ensembles with valence quark masses from $\sim m_s/8$ -- $m_s$.   In order to keep finite volume effects under control, we restrict the combination $a m_\pi L \gtapprox 4$;  our lightest pion is $\sim 280$ MeV.  Unlike in other mixed staggered sea, domain-wall valence simulations, we do not tune our valence-valence pion masses to any particular value.  Because there is no notion of full QCD at nonzero lattice spacing in the mixed action theory, we generate many partially quenched data points, and we use the appropriate partially quenched, mixed action $\chi$PT to extrapolate our lattice data.  Our analysis is still in progress, and in this section we present only preliminary data and chiral fits.

 A necessary prerequisite for the calculation of $B_K$ is the determination of the pseudoscalar decay constants $f_\pi$ and $f_K$.  Because they are known quantities, they provide a test of the lattice methodology, and allow one to demonstrate an understanding of and control over systematic errors.  In order to suppress contamination from pions circling the lattice in the time direction, we use symmetric and antisymmetric linear combinations of periodic and antiperiodic quark propagators in our correlation functions.  We then extract the decay constant using the axial Ward identity:
 \begin{equation}
 	f_P = \frac{A_{WP}}{\sqrt{A_{WW}}} \frac{\sqrt{2} (m_x + m_y + 2 m_{res})}{m_\pi^{3/2}} ,
\label{eq:fpi} \end{equation}
 where $A_{WW}$, $A_{WP}$, and $m_\pi$ come from a simultaneous fit of wall-wall and wall-point correlators. 

\begin{figure}
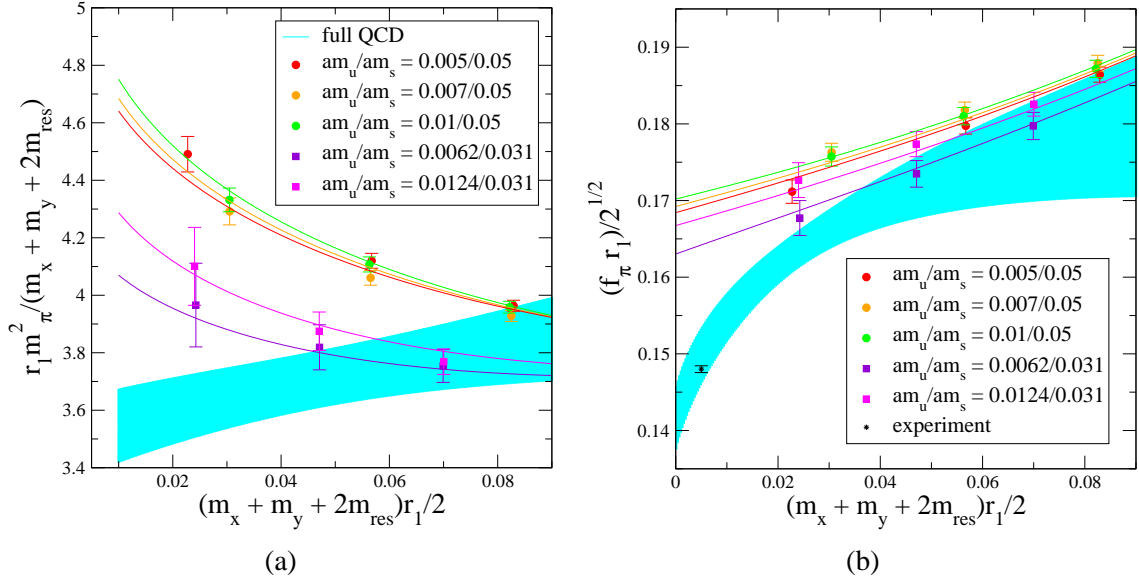

\begin{center}
\begin{tabular}{c c}
\includegraphics[scale=0.405,angle=0]{mpi_5ens.eps}  & \includegraphics[scale=0.405,angle=0]{fpi_5ens.eps}  \\
(a) & (b) 
\end{tabular}
\end{center}
\vspace{-3mm}
\caption{``Pion'' mass-squared and decay constant versus average quark mass.  The circles are coarse data points and the squares are fine data points.  The cyan band is the continuum full QCD curve (with statistical errors) that results from fitting the lattice data to the appropriate mixed action $\chi$PT expression; fit (a) has a $\textrm{CL}=0.34$ and fit (b) has a $\textrm{CL}=0.83$.  Note that the band in plot (b) contains the experimental value of $f_\pi$. }
\label{fig:fpi}
\end{figure}

Figure~\ref{fig:fpi} shows preliminary chiral and continuum extrapolations of the ``pion'' mass and decay constant data (i.e., $m_x = m_y$) using the approach of the MILC collaboration described in Ref.~\cite{CB_cuts}.  
They are each simultaneous fits to the coarse and fine data using the next-to-leading order (NLO) partially quenched, mixed action $\chi$PT expression for $m_\pi$ or $f_\pi$, plus additional NNLO analytic terms.  In order to achieve good correlated fits without needing NNLO logarithm terms (the full set of which are not known for the mixed action theory), we omit valence points with $m_\pi \gtapprox 500$~MeV and all points on the heaviest ($m_l / m_s = 0.4$) coarse ensemble.  We also correct the data by the known 1-loop finite volume effects.  Note that, with these judicious cuts, we have 30 data points for only 7 fit parameters, and do not need to introduce Bayesian constraints.     

Although the fitting procedure described above successfully reproduces the experimental value of $f_\pi$, it is not the only reasonable possibility, and  various physically motivated approaches are used in the literature.  One can, in principle, continue to remove heavy masses until unadulterated NLO $\chi$PT gives an acceptable fit.  Typical lattice simulations, however, only have one or two data points in this regime, making this method largely impractical.  Furthermore, one cannot do this for quantities involving $m_{strange}$.  One can instead include heavier data points at the cost of introducing terms of higher-order in $\chi$PT.   Unfortunately, however, the NNLO chiral logarithms are known only for the pseudoscalar masses and decay constants and only in the continuum~\cite{Bijnens}.  It is likely that the inclusion of only NNLO analytic terms does not significantly impact extrapolated results for physical quantities~\cite{Sharpe}, but, of course, it would be preferable to include the full NNLO expression.  Finally, use of $SU(2)$ $\chi$PT provides a valuable crosscheck on the determination of $m_\pi$ and $f_\pi$, but cannot be used to determine quantities involving valence strange quarks (such as $f_K$ and $B_K$) without further assumptions.  One can either match an $SU(2)$ $\chi$PT fit of light quark masses onto another fit of data near $m_{strange}$, which requires an additional fit ansatz, or one can treat the strange quark as a heavy quark within $SU(2)$ heavy meson $\chi$PT, but this relies on the assumption that $m_{strange}$ is heavy compared to $\Lambda_\textrm{QCD}$.  Ultimately, the degree to which these fit choices change results for extrapolated physical quantities must be reflected in the systematic errors.  

Although we do not yet have sub-percent statistical error bars like the MILC Collaboration, we can still begin to distinguish between fit ans\"atze that work and those that do not, and hence draw interesting conclusions about the application of $\chi$PT to lattice QCD.  In order to assess whether or not we have sufficient data in the chiral regime to allow the use of $\chi$PT, we must first look for chiral logarithms in the pseudoscalar data.  We observe the appropriate curvature in the quantity $m_\pi^2 / m_q$, which is expected to have large partially quenched chiral logarithms.  We also find that, in order to get a fit of the pion decay constant data with an acceptable confidence level, we must include the NLO chiral logarithms.  Thus we have both qualitative and numerical evidence for the presence of chiral logarithms in our data.  Not only must we include chiral logarithms for good fits, but in fact we need the \emph{correct} mixed action chiral logarithms.  We can no longer fit the decay constant data if we set the taste-breaking mass-splittings $\Delta_I$, $\Delta_\textrm{mix} \to 0$.  Thus we conclude that it is important to use the $\chi$PT expression appropriate to the specific lattice action used in simulations.\footnote{Another quantity where inclusion of the correct lattice discretization effects is necessary is the isovector scalar correlator.  We cannot describe the behavior of our $a_0$ correlator data~\cite{scalar} if we set $\Delta_\textrm{mix}\to 0$ in the mixed action $\chi$PT expression for the disconnected bubble contribution~\cite{Sasa}.}  We also implicitly observe finite-volume effects in our data.   Despite our reasonably conservative choice to keep the quantity $m_\pi L \gtapprox 4$, the 1-loop finite-volume corrections to our $m_\pi^2$ data are as large as $\sim3\%$ due to enhancements from partial quenching, and we cannot get an acceptable fit to our pion mass data without their inclusion.  Thus we are currently performing a finite-volume study on one of the coarse ensembles.  Finally, our decay constant fit clearly resolves a large $\CO(a^2)$ analytic term that causes the coarse data to lie $\sim10\%$ above the continuum full QCD curve.  Because generic discretization effects will likely be of this size for all simulations not using $\CO(\alpha_S a^2)$ improved lattice actions, this underscores the importance of having multiple lattice spacings.  

\begin{figure}
\begin{center}
\includegraphics[scale=0.43,angle=0]{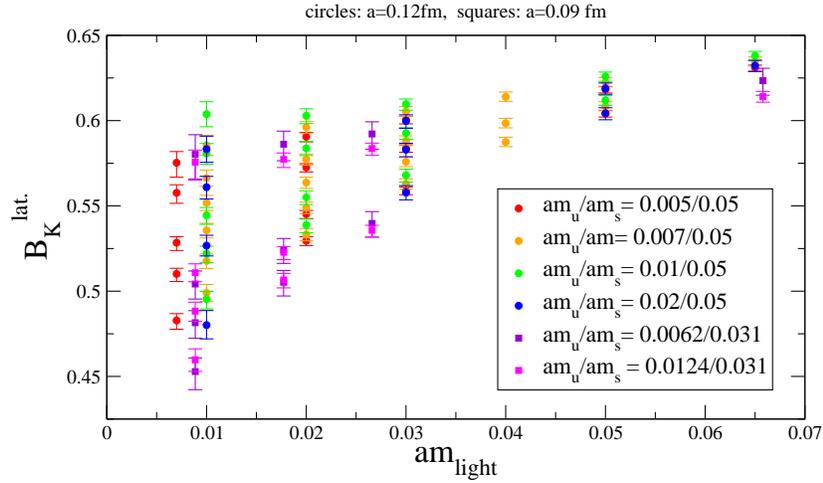} \end{center}
\vspace{-3mm}
\caption{Lattice $B_K$ versus light quark mass in ``kaon''.}
\label{fig:BK}
\end{figure}

We calculate the $K^0-\overline{K^0}$ matrix element using Coulomb gauge-fixed wall-source propagators, and then determine $B_K$ with the following ratio:
 \begin{equation}
 	B_K^{lat.} = \frac{V}{8/3} \frac{C_{WPW}^{P\CO P} (t_{source},t,t_{sink})}{C_{WP}^{PA}(t_{source},t) C_{WP}^{PA}(t_{sink},t)} ,
 \end{equation}
 where, in the upper labels, ``$P$" indicates the pseudoscalar operator, ``$\CO$" indicates the 4-quark operator, and ``$A$" indicates the axial operator.  Figure~\ref{fig:BK} shows the resulting preliminary data.

\section{Summary and future plans}

$B_K$ must be known to an accuracy of 5\% or better in order to have phenomenological impact.  We have already calculated the expression for $B_K$ in mixed action $\chi$PT and shown that, at NLO, it has only one more low-energy constant than in the continuum~\cite{mixbk}.  We have determined the sizes of $m_{res}$ and  $\Delta_\textrm{mix}$ on the MILC lattices and shown that both the amount of chiral symmetry breaking and taste-symmetry breaking are under control.  We have generated correlation functions necessary for the calculation of pseudoscalar masses, decay constants and $B_K$ at several sea quark masses and lattice spacings, and have begun chiral and continuum extrapolations. In the near future we plan to finish the nonperturbative renormalization and determine the domain-wall valence strange quark mass.  Our preliminary results are promising. The errors in our $B_K$ data range from 0.5--2\% and will continue to improve, and our preliminary determination of $f_\pi$ has a $2.5\%$ statistical uncertainty.  This suggests that the mixed action method will allow a precise determination of $B_K$.
  
\section*{Acknowledgements}

Our calculations were carried out on USQCD computing resources.  We thank the MILC Collaboration for use of their lattices and Andreas Kronfeld and Steve Sharpe for helpful comments.



\begin{thebibliography}{99}

\bibitem{JLQCD_BK}
  S.~Aoki {\it et al.}  [JLQCD Collaboration],
  Phys.\ Rev.\ Lett.\  {\bf 80}, 5271 (1998)
  [arXiv:hep-lat/9710073].
  
\bibitem{HPQCD_BK}
  E.~Gamiz {\it et al.} [HPQCD Collaboration],
  Phys.\ Rev.\  D {\bf 73}, 114502 (2006)
  [arXiv:hep-lat/0603023].
  
\bibitem{RBC_BK}
  D.~J.~Antonio {\it et al.}  [RBC and UKQCD Collaborations],
  arXiv:hep-ph/0702042.
  
\bibitem{bigMILC}
  C.~Aubin {\it et al.}  [MILC Collaboration],
  Phys.\ Rev.\  D {\bf 70}, 114501 (2004)
  [arXiv:hep-lat/0407028].
  
\bibitem{Chroma}
  R.~G.~Edwards and B.~Jo\'o,
  Nucl.\ Phys.\ Proc.\ Suppl.\  {\bf 140}, 832 (2005)
  [arXiv:hep-lat/0409003].
  
\bibitem{NPR}
  G.~Martinelli {\it et al.},
  Nucl.\ Phys.\  B {\bf 445}, 81 (1995)
  [arXiv:hep-lat/9411010].
  
\bibitem{Susskind}
  L.~Susskind,
  Phys.\ Rev.\ D {\bf16}, 10 (1977).
  
\bibitem{LeeSharpe}
  W.~J.~Lee and S.~R.~Sharpe,
  Phys.\ Rev.\  D {\bf 60}, 114503 (1999)
  [arXiv:hep-lat/9905023].

\bibitem{DWF}
  D.~B.~Kaplan,
  Phys.\ Lett.\  B {\bf 288}, 342 (1992)
  [arXiv:hep-lat/9206013];\\
  Y.~Shamir,
  Nucl.\ Phys.\  B {\bf 406}, 90 (1993)
  [arXiv:hep-lat/9303005].
    
\bibitem{LHPC}
  D.~B.~Renner {\it et al.}  [LHP Collaboration],
  Nucl.\ Phys.\ Proc.\ Suppl.\  {\bf 140}, 255 (2005)
  [arXiv:hep-lat/0409130].

\bibitem{Bar}
  O.~B\"ar {\it et al.},
  Phys.\ Rev.\  D {\bf 72}, 054502 (2005)
  [arXiv:hep-lat/0503009].
    
\bibitem{MixMeson}
  K.~Orginos and A.~Walker-Loud,
  arXiv:0705.0572 [hep-lat].
  
\bibitem{CB_cuts}
  C.~Bernard {\it et al.} [MILC Collaboration], 
  \emph{these proceedings}.
  
\bibitem{Bijnens}
   J.~Bijnens, N.~Danielsson and T.~A.~Lahde,
  Phys.\ Rev.\  D {\bf 73}, 074509 (2006)
  [arXiv:hep-lat/0602003].
  
\bibitem{Sharpe}
  S.~R.~Sharpe,
  arXiv:hep-lat/0607016.
  
 \bibitem{scalar}
   C.~Aubin, J.~Laiho and R.~S.~Van de Water,
   \emph{these proceedings}.
 
 \bibitem{Sasa}
  S.~Prelovsek,
  Phys.\ Rev.\  D {\bf 73}, 014506 (2006)
  [arXiv:hep-lat/0510080].

\bibitem{mixbk}
  C.~Aubin, J.~Laiho and R.~S.~Van de Water,
  Phys.\ Rev.\  D {\bf 75}, 034502 (2007)
  [arXiv:hep-lat/0609009].

\end{thebibliography}
\end{document}